\documentclass[11pt]{article}
\usepackage{jhep}
\usepackage[T1]{fontenc}
\usepackage{bm}
\usepackage{graphicx}
\usepackage{float}
\usepackage{caption}
\usepackage{subcaption}
\usepackage[normalem]{ulem}
\usepackage{gensymb}
\usepackage[utf8]{inputenc}
\usepackage{bbold}
\usepackage{soul}
\usepackage{slashed}
\usepackage[dvipsnames]{xcolor}
\usepackage{cleveref}
\hypersetup{colorlinks=false}

\preprint{UCI-HEP-TR-2020-02}

\newcommand{\MeV}{\text{MeV}}

\newcommand{\munu}{{\mu \nu}}
\newcommand{\Lnew}{ \Lambda_{\rm QCD}}
\newcommand{\LSM}{ \Lambda_{\rm QCD}^{\rm SM}}
\newcommand{\tr}{\mathrm{tr}}
\newcommand{\Lagr}{\mathcal{L}}

\title{Dark Matter Freeze Out during an Early Cosmological Period of QCD Confinement}
\author{Dillon Berger,}
\author{Seyda Ipek,}
\author{Tim M.P. Tait,}
\author{Michael Waterbury}

\affiliation{Department of Physics and Astronomy, University of California, Irvine, CA 92697 USA}
\emailAdd{bergerdt@uci.edu}
\emailAdd{sipek@uci.edu}
\emailAdd{ttait@uci.edu}
\emailAdd{mwaterbu@uci.edu }

\date{\today}
\abstract{Standard lore states that there is tension between the need to accommodate the relic density
of a weakly interacting massive particle and direct searches for dark matter.  However, the estimation of
the relic density rests on an extrapolation of the cosmology of the early Universe to the time of freeze out,
untethered by observations.  We explore a nonstandard cosmology in which the strong coupling constant
evolves in the early Universe, triggering an early period of QCD confinement at the time of freeze out.
We find that depending on the nature of the interactions between the dark matter and the Standard Model,
freeze out during an early period of confinement can lead to drastically different expectations for the relic
density, allowing for regions of parameter space which
realize the correct abundance but would otherwise be excluded by direct searches.
}

\begin{document}
\maketitle

\section{Introduction}

The identity of the dark matter, necessary to explain a host of cosmological observations, is among the most pressing
questions confronting particle physics today.  The Standard Model (SM) contains no suitable fields to play the role of dark matter,
and understanding how it must be amended to describe dark matter will inevitably provide important insights
into the theory of fundamental particles and interactions.  There are a plethora of theoretical ideas as to how to incorporate
dark matter, and exploring how to test them is a major area of activity in particle experiment.

Among the various candidates, the class of weakly interacting massive particles (WIMPs) remains extremely attractive,
largely driven by the appealing opportunity to explain their relic density based on the strength
of their interactions with the SM.  Provided their
interactions are roughly similar to the electroweak couplings, WIMPs are expected to
initially be in chemical equilibrium with the SM plasma at early times, but to fall out of equilibrium
when the temperature of the Universe falls below $T \sim m_\chi / 20$, where $m_\chi$ is the
mass of the WIMP.  Provided the mass and cross section for annihilation into the SM are
correlated appropriately \cite{Feng:2008ya}, the observed cosmological abundance is relatively easily realized.

Vanilla theories of WIMPs are challenged by the null results from direct searches for dark matter scattering with heavy
nuclei \cite{Aprile:2018dbl}.
For many generic models of WIMP interactions with the SM, these searches exclude the required annihilation cross
section for masses $1~{\rm GeV}\lesssim m_\chi\lesssim 10^4~$GeV.  While it is possible to engineer
interactions that allow for large annihilation while suppressing scattering (see
\cite{TuckerSmith:2001hy,Cirelli:2005uq,Goodman:2010ku,Freytsis:2010ne,Boehm:2014hva,Ipek:2014gua,Abdullah:2014lla} for
a few examples), such limits, together with those derived from the null observations of WIMP annihilation
products \cite{Ahnen:2016qkx} and/or
production at colliders \cite{Beltran:2008xg,Aaboud:2017phn,Sirunyan:2017hci}, suggest that
either Nature has been unkind in choosing which model of WIMPs to realize, or there is tension between realizing the
observed relic density and the limits from experimental searches for WIMPs.

A key assumption under-pinning the mapping of the relic density to WIMP searches today is that the cosmological
history of the Universe can be reliably extrapolated back to the time of freeze out.
The standard picture extrapolates based on a theory containing the SM plus dark matter (and dark energy), with
no other significant ingredients.
The success of Big Bang Nucleosynthesis (BBN)
in explaining the primordial abundances of the light elements could be taken as an argument
that it is unlikely that cosmology has been very significantly altered at temperatures lower than $\sim 10$~MeV,
but this is far below the typical freeze-out temperature of a weak scale mass WIMP, which is more typically in
the 5-100 GeV range.
Indeed, it has been shown that an early period of matter domination \cite{Hamdan:2017psw} or late entropy production \cite{Gelmini:2006pw}
can alter the relic abundance for fixed WIMP model parameters, leading to substantially different mapping
between the observed abundance and the expectations of direct searches.

In this article, we explore a different kind of nonstandard cosmology, in which the strong interaction described by
Quantum Chromodynamics (QCD) undergoes an early phase of confinement, based on promoting the strong coupling
$\alpha_s$ to a field, whose potential receives thermal corrections which cause it to take larger values at early times,
relaxing to the canonical size some time before BBN \cite{Ipek:2018lhm,Croon:2019ugf}.
If the dark matter freeze out occurs during a period in which
$\alpha_s$ is larger such that QCD is confined, the degrees of freedom of the Universe are radically different
from the naive extrapolation,
being composed largely of mesons and baryons rather than quarks and gluons.  Similarly, the interactions of the
dark matter with the hadrons are scaled up by the larger QCD scale, $\Lambda_{\rm QCD}$, leading to a very different annihilation
cross section at the time of freeze-out than during the epoch in which experimental bounds are operative.  We find that
depending on the underlying form of the dark matter interactions with quarks, radical departures from the expected
relic density are possible.

This article is organized as follows.  In Section~\ref{sec:confinement}, we review the construction of a Universe in which
$\alpha_s$ varies with temperature.
In Section~\ref{sec:DMchiral} we discuss the chiral perturbation theory which describes the mesons and their interactions
with the dark matter during the period of early confinement, and
in Section~\ref{sec:freezeout}, we examine the relic density under different assumptions
concerning $\alpha_s$ at the time of freeze-out, and contrast with experimental constraints derived today.  We reserve
Section~\ref{sec:conclusions} for our conclusions and outlook.

\section{Early QCD Confinement}
\label{sec:confinement}

Following reference~\cite{Ipek:2018lhm}, we modify the gluon kinetic term in the SM Lagrangian to:
\begin{align}
 -\frac{1}{4 g^2_{s0}}  G^a_\munu G_a^\munu ~~~\Rightarrow~~~
 -\frac14 \left( \frac{1}{g^2_{s0}} + \frac{S}{M_\ast} \right) G^a_\munu G_a^\munu~,
 \label{eq:Lag}
\end{align}
where $G^a_\munu$ is the gluon field strength,
$S$ is a gauge singlet real scalar field, and $g_{s0}$ represents
(after rescaling the kinetic term to canonical normalization) the
$SU(3)$ gauge coupling in the absence of a vacuum expectation value (VEV) for $S$.
$M_\ast$ is a parameter with dimensions of energy which parameterizes a non-renormalizable interaction
between $S$ and the gluons. It could represent the fluctuations of a radion or dilaton field, or by
integrating out heavy vector-like $SU(3)$-charged particles which also couple to the scalar field $S$.
In the latter case, the scale of the interaction is related to the mass of the new $SU(3)$-charged
particles via $M_\ast \sim 4\pi M_Q/n_Q y_Q\alpha_s$,
where $n_Q$ is the number of $SU(3)$-charged fermions with mass $M_Q$ and Yukawa coupling $y_Q$.

Engineering an early period of confinement, followed by subsequent deconfinement
and return to a SM-like value of $\alpha_s$ before BBN
imposes constraints on the potential for $S$, and its interactions with other fields (which determine the thermal
corrections to its potential) \cite{Ipek:2018lhm}.  Generally, mixed potential terms containing the SM Higgs doublet are present,
and these may play an important role in the thermal history \cite{Croon:2019ugf}.  In this work,
we remain agnostic concerning the specific dynamics which implement the shift in $v_S$ leading to early
confinement, and we assume that the terms mixing the $S$ with the SM Higgs are small enough so as to
be safely neglected.

\begin{figure}
    \centering
    \includegraphics[width=.45\textwidth]{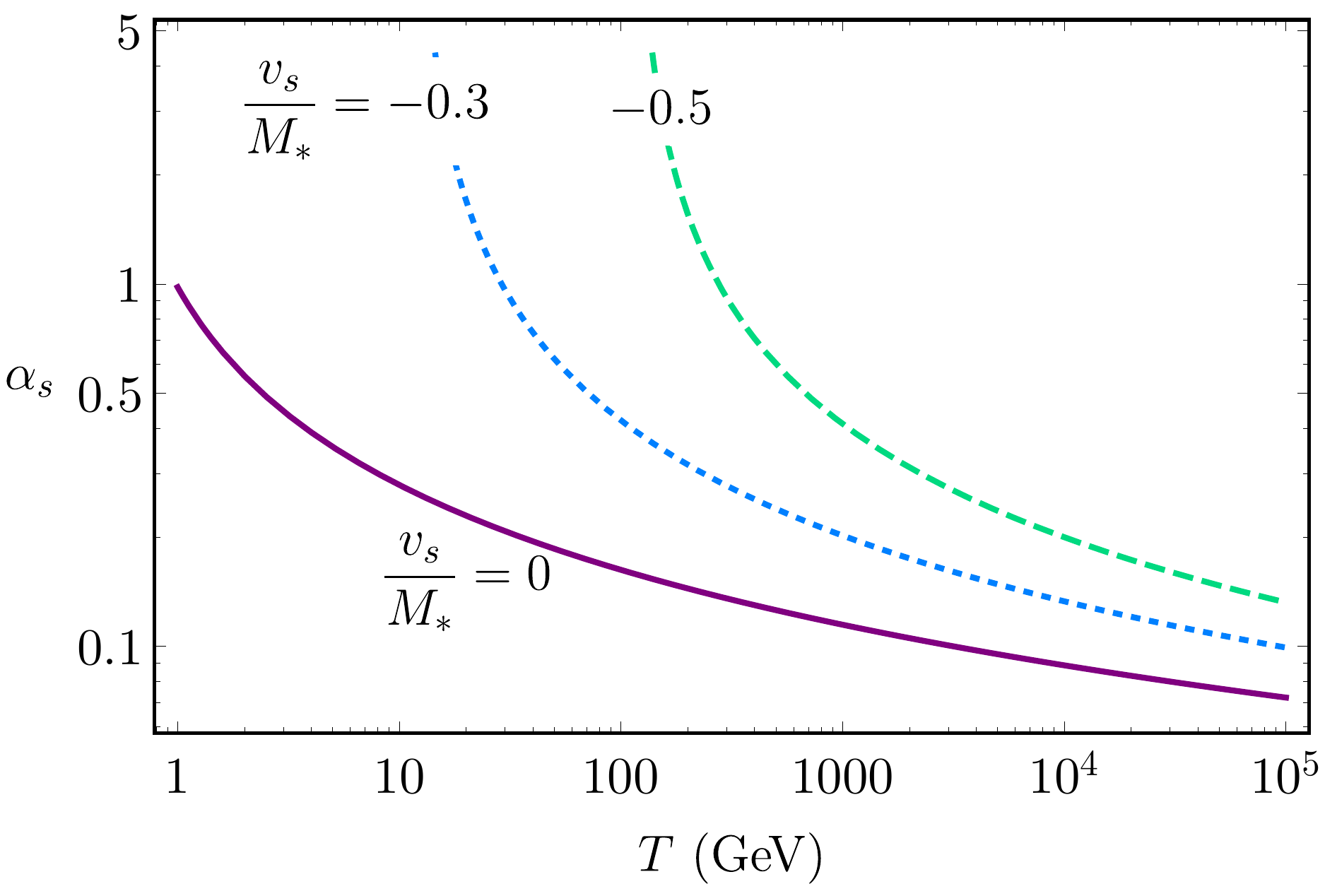}~~~~~~
    \includegraphics[width=.47\textwidth]{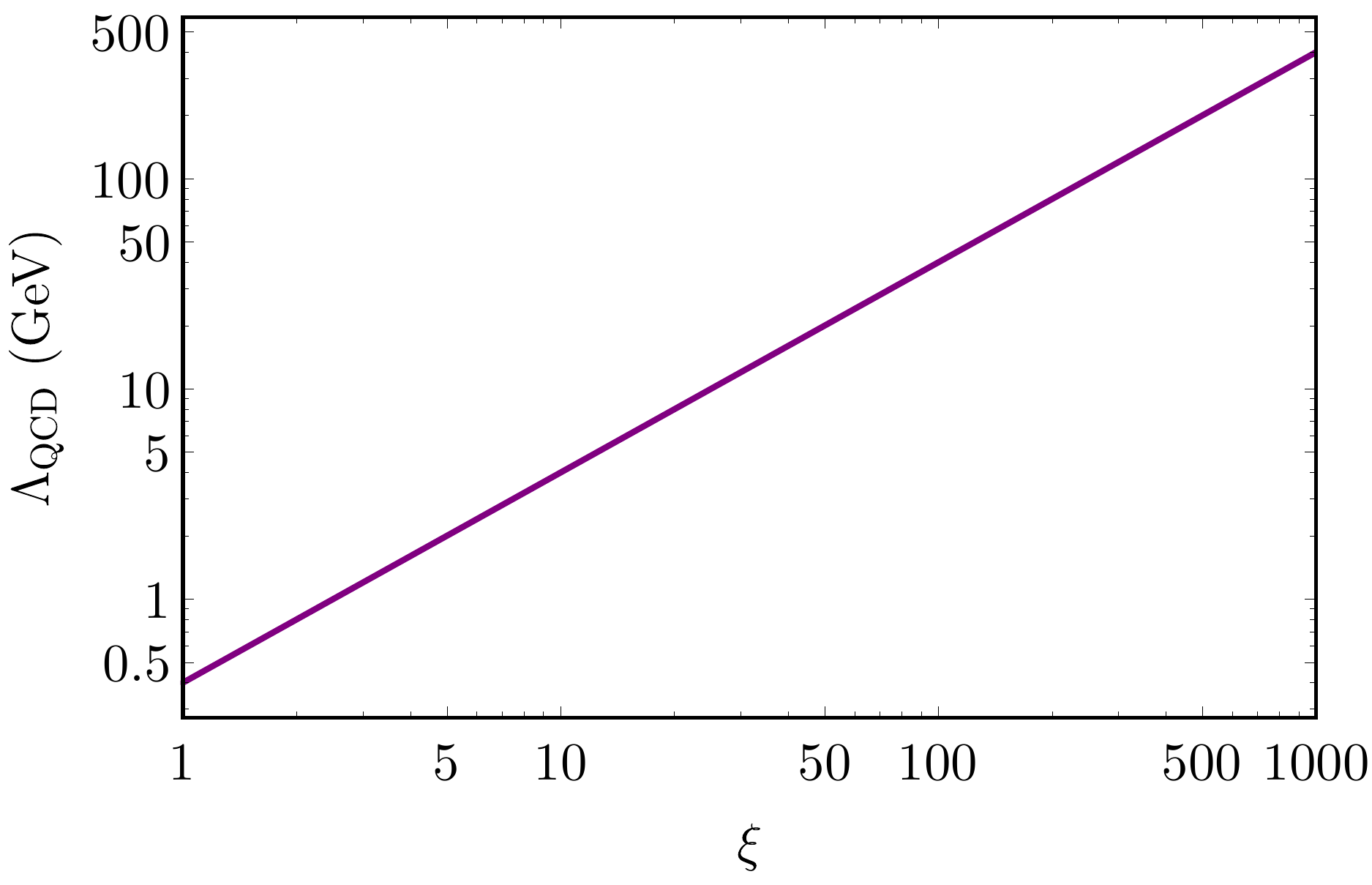}
    \caption{\textbf{Left panel:} Evolution of the strong coupling constant with temperature in the early Universe for
    three different values of $v_s / M_{\ast}$.
    Confinement takes place at temperatures for which $\alpha_s \gg 1$.
    \textbf{Right panel:} The scale of QCD confinement, $\Lnew$, as a function of the parameter
    $\xi = \exp(24\pi^2/(2N_f -33)v_S/M_\ast)$.}
    \label{fig:alphasT}
\end{figure}

A VEV for $S$ generates a non-decoupling correction to the effective strong coupling constant through the
dimension-5 interaction in \Cref{eq:Lag}, which for negative $v_s$ strengthens the effective coupling strength.
At one loop and at scale $\mu$, the effective strong coupling is
\begin{align}
    \frac{1}{\alpha_s(\mu,v_s)}= \frac{33-2N_f}{12\pi}\,{\rm ln}\left(\frac{\mu^2}{\Lambda_0^2}\right)+4\pi\frac{v_s}{M_\ast},
\end{align}
where $N_f$ is the number of active quark flavors at the scale $\mu \sim T$.
\Cref{fig:alphasT} shows the effective coupling as a function of temperature.
QCD confinement occurs at a temperature $T_c \simeq \Lambda_{\rm QCD}$, where
\begin{align}
    \Lnew(v_s) = \LSM\, e^{\frac{24 \pi^2}{2 N_f - 33} \frac{v_s}{M_\ast}}. \label{eq:LQCD}
\end{align}
Here, $\LSM \simeq 400~$MeV is the SM value of the QCD confinement scale; we adjust
$g_{s0}$ such that it is realized for $v_s =0$.

At scales below confinement, the relevant degrees of freedom are mesons, whose
dynamics are described by chiral perturbation theory, the effective field theory
of which is parameterized by coefficients which
depend on $\Lnew$.  We find it convenient to parameterize the physics in terms of the ratio of $\Lnew$ to
$\LSM$,
\begin{align}
\xi \equiv \frac{\Lnew}{\LSM}\simeq \exp\left({\frac{24 \pi^2}{2 N_f - 33} \frac{v_s}{M_\ast}}\right).
\end{align}
The parameter $\xi$ is typically sufficient to completely describe the physics of dark matter interactions during
the period of early confinement.

\section{Dark Matter Interactions and Chiral Perturbation Theory}
\label{sec:DMchiral}

The dynamics of the scenario we study are encoded in the Lagrangian:
\begin{align}
\Lagr \supset -\frac14 \left( \frac{1}{g^2_{s0}} + \frac{S}{M_\ast} \right) G^a_\munu G_a^\munu +
\sum_q \left\{ i \bar{q}\slashed{D}q - y_q\,h \bar{q}_L  q_R  +{\rm H.c.} \right\} + \Lagr_\chi\, ,
\end{align}
where $\Lagr_\chi$ describes the dark matter and its interactions. We introduce a SM-singlet Dirac fermion field $\chi$ to represent the dark matter, and couple it to quarks,
\begin{eqnarray}
{\cal L}_\chi = i \bar{\chi} \gamma^\mu \partial_\mu \chi - m_\chi \bar{\chi} \chi
+ \sum_{\bar i,j} \left\{
\frac{\beta_{ij}}{M_S^2} \bar{\chi} \chi ~\bar{q}_i  q_j
+ \frac{\lambda_{ij}}{M_V^2} \bar{\chi} \gamma^\mu \chi ~ \bar{q}_i \gamma_\mu  q_j
\right\} \, ,
\label{eq:LDM}
\end{eqnarray}
where the couplings $\beta_{i j} / M_S^2$ and $\lambda_{i j} / M_V^2$ represent
operators left behind by integrating out states with masses $\gg m_\chi$.
Generically, one would also expect there to be interactions with the leptons or the Higgs doublet.  We assume for simplicity that such interactions are subdominant if present.

In the case of the scalar interactions, $S$ itself could act as the mediator,
provided it has direct coupling to the dark matter.  In that case, UV-completing will require additional states
to provide a renormalizable portal to $h \bar{q} q$, and the dimension six interaction written here will descend from
a dimension seven operator after the SM Higgs gets its VEV.  The vector interactions could represent a $Z^\prime$
from an additional U(1) gauge symmetry that couples to both quarks and dark matter.  We will
consider cases in which either scalar or vector interactions dominate over the other one. We follow the
guidance of minimal flavor violation \cite{DAmbrosio:2002vsn}
in choosing the couplings such that
\begin{eqnarray}
\beta_{ij} \equiv \pm \delta_{ij}  \frac{y_i}{y_u}~,
\label{eq:defbeta}
\end{eqnarray}
which is normalized to the coupling to up quarks, and with an over-all factor absorbed into $M_S^2$.
The possibility of choosing either sign for $\beta$ will play an important role, described in \ref{eqn:mshift} below.

The vector couplings $\lambda_{ij}$ are diagonal and
have equal values for the up-type quarks, and equal (but different from the up-type) values for the down-type quarks,
\begin{equation}
    \lambda_{ i j} \equiv \begin{cases}
        \delta_{i j}, & j={u,c,t} \\
        (1+\alpha)\delta_{i j}, & j={d,s,b}~,
    \end{cases}
    \label{eq:deflambda}
\end{equation}
where $\alpha$ determines the difference between up- and down-type couplings. When $\alpha = 0$, the vector coupling assigns charges equivalent to baryon number, and the mesons decouple from the dark matter.

During early confinement, the Universe looks very different from the standard cosmological picture based on the SM extrapolation.
(Massless) quark and gluon degrees of freedom are replaced by mesons and baryons, and chiral symmetry breaking
induces a tadpole for the Higgs which is relevant for the evolution of its VEV. In order to determine how dark matter interactions are affected by
this early cosmological period of QCD confinement, we first give a description of this era in terms of chiral perturbation theory.

\subsection{Chiral Perturbation Theory}
\label{sec:chipt}

In the limit $\Lnew \gg m_t$, the QCD sector of the Lagrangian for quarks,
\begin{align}
\Lagr \supset
\sum_q \left\{ i \bar{q}\slashed{D}q - y_q\,h \bar{q}_L  q_R  +{\rm H.c.} \right\}
\end{align}
(where $h$ is the SM Higgs radial mode)
possesses an approximate global SU(6)$_L \times$ SU(6)$_R$ chiral symmetry, which is softly broken by the Yukawa
interactions.  We work in the basis in which the $y_q$'s are diagonal, for which all flavor-changing processes reside
in the electroweak interactions.  Non-perturbative QCD is expected to break
SU(6)$_L \times$ SU(6)$_R \rightarrow SU(6)_V$ to the diagonal subgroup, resulting in
$6^2 - 1 = 35$ pions as pseudo-Nambu-Goldstone bosons.

\begin{figure}[t]
\centering
\includegraphics[scale=.6]{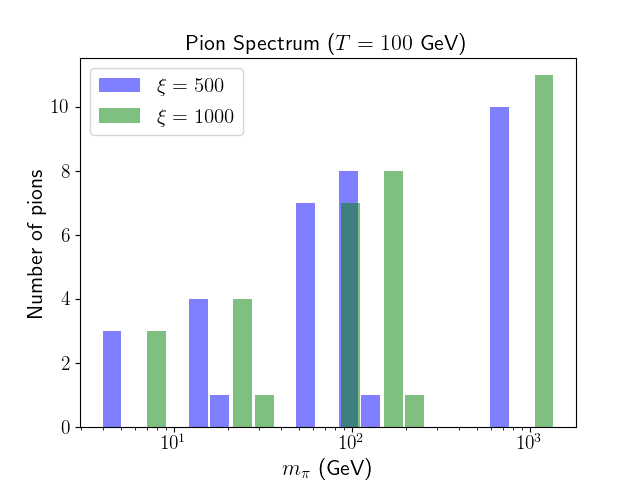}
\caption{Spectrum of pion masses for two choices of $\xi$, with $v_h$ corresponding to the Higgs VEV at $T=100$~GeV.
}\label{fig:pispectrum}
\end{figure}

At scales below $\Lnew$, the pions are described by a nonlinear sigma model
built out of $U(x) \equiv \exp \left( i2T^a \pi^a(x) /f_\pi \right)$, where $T^a$ are the $SU(6)$ generators. The leading terms
in the chiral Lagrangian (neglecting electroweak interactions) are
\begin{align}
\Lagr_{\rm ch} = \frac{f_\pi^2}{4} ~\tr(|D_\mu U|^2) + \kappa\,~ \tr(U M_q^\dagger + M_q  U^\dagger)~, \label{eq:Lchi}
\end{align}
where $f_\pi$ is the pion decay constant and $\kappa$ is a constant with mass dimension 3,
both of which represent the strong dynamics. The generators are normalized such that $\tr[T^aT^b]=\delta^{ab}/2$, leaving the $\pi^a$ canonically normalized.
The mass matrix $M_q$ is a spurion representing the explicit SU(6)$_L \times$ SU(6)$_R$ breaking from
the Yukawa interactions,
\begin{align}
M_q = \frac{1}{\sqrt{2}} h ~\mathrm{Diag}(y_u, y_d, y_s, y_c, y_b, y_t)~.
\end{align}
Expanding the field $U$ in \Cref{eq:Lchi} to second order in $\pi / f_\pi$ results in pion
mass terms and a tadpole for the Higgs:
\begin{align}
\Lagr_{\rm ch} \supset \sqrt{2}\kappa\, y_t\, h- \frac{\kappa}{f_\pi^2}\, \tr[\{T^a,T^b\}M]\,{\pi}^a {\pi}^b~,
\end{align}
both of which are controlled by $\kappa$. (In the tadpole term we keep only the top Yukawa as the contributions
from light quarks are typically negligible.)
We match $f_\pi$ and $\kappa$  to the SM pion mass,
$m_{\pi 0} = 135~$MeV, and decay constant, $f_{\pi 0}=94~$MeV
at $\xi=1$ and $v_h = v_h^0$, where $v_h^0=246~$GeV is the zero temperature SM Higgs VEV.
Naive dimensional analysis provides the scaling for other values of $\xi$
(for which the tadpole implies there will typically be a different $v_h$):
\begin{align}
\kappa \simeq  (220~\MeV)^3\,\xi^3~,~~~~ f_\pi \simeq 94~\MeV\, \xi, ~~~~ m_\pi^2 \simeq m_{\pi 0}^2\, \xi\, v_h/v_h^0,
\end{align}
The resulting pion mass matrix is diagonalized numerically to determine the spectrum of mesons in the
mass basis.  Example spectra at $T= 100$~GeV for two different choices of $\xi$ are shown in Figure~\ref{fig:pispectrum}.

\subsection{Finite Temperature Higgs Potential}
\label{sec:finiteThiggs}

As shown above, a cosmological era of early QCD confinement induces a tadpole for the Higgs radial mode $h$.
If $\Lnew$ is comparable in size to the weak scale, this tadpole can deform the Higgs potential by a relevant amount
during the epoch of confinement.  In addition, the plasma contains mesons (rather than quarks), which modifies the
thermal corrections to the Higgs potential from the SM fermions.

We determine the Higgs VEV  as a function of temperature
by finding the global minimum of the finite-temperature Higgs potential.
We assume that interaction terms between the Higgs and $S$ are small enough to be neglected.
We focus on a cosmological history where $\Lnew > T_{\rm EW}\sim 150~$GeV, which requires $\xi\gtrsim 300$.
We further assume that the $S$ potential is such that there is a lower temperature $T_d$
(which we treat as a free parameter) at which $\Lnew$ returns to
$\LSM$, implying that QCD deconfines and the subsequent evolution of the Universe is SM-like.

Under these assumptions, the finite temperature potential for the Higgs, $V(h, T)$ consists of the
tree level SM potential,
\begin{align}
V_0(h)=-\frac{1}{2}\mu^2h^2 + \frac{\lambda}{4} h^4~,
\end{align}
whose parameters are adjusted to match the zero temperature VEV $v_h^0 = 246$~GeV and Higgs mass
$m_h \simeq 126$~GeV. In three different temperature regimes, the form of the finite temperature corrections is given as
\begin{align}
V(h,T) =\left\{ \begin{array}{lc}
 {\displaystyle V_0(h) + \frac{T^{4}}{2 \pi^{2}} \sum_{i=h, W, Z,t}\!\!\!\! (-1)^F n_{i} J_{B/F}\left[m_{i}^{2} / T^{2}\right] }&
 ~~~~~~~~~ \left( T>\Lnew \right)\\
{\displaystyle V_0(h) - \sqrt{2}\kappa y_t h+  \frac{T^{4}}{2 \pi^{2}} \sum_{i=h, W, Z, \pi^a} \!\!\!\!\!\!\!\!n_{i} J_{B}\left[m_{i}^{2} / T^{2}\right]} & ~~~~~~~~~ \left(T_d<T< \Lnew \right),\\
{\displaystyle V_0(h)  + \frac{T^{4}}{2 \pi^{2}} \sum_{i=h, W, Z, t}\!\!\!\! (-1)^F n_{i}  J_{B/F}\left[m_{i}^{2} / T^{2}\right]} &
~~~~~~~~~ \left(T<T_d \right)~,
\end{array}\right.
\label{Veff}
\end{align}
where $F=0/1$ for bosons/fermions and $n_i$ counts degrees of freedom:
$n_h=n_\pi=1$, $n_{W}=6$, $n_Z =3$, and $n_t=12$. The functions $J_{B/F}$ are the bosonic/fermionic thermal functions,
\begin{align}
J_{B/F}\left[m_i^2/T^2\right] &= \int_{0}^\infty x^2\log\left(1 - (-1)^F e^{-\sqrt{x^2 + m_i^2/T^2}} \right) \label{thermal_func}
\end{align}
and $m_i^2(h)$ are the field dependent masses,
\begin{align}
m_h^2 &= -\mu^2 + 3\lambda h^2 , ~~~~~~
m_W^2 = \frac{g_W^2}{4} h^2, ~~~~~~
m_Z^2 = \frac{g_W^2}{4\cos^2(\theta_w)} h^2, ~~~~~~
m_t^2 = \frac{y_t^2}{2} h^2.
\end{align}

We make use of the high temperature expansions of the thermal functions, which are given as

\begin{align}
J_{B}\left(m^{2}(h) / T^{2}\right)&=-\frac{\pi^{4}}{45}+\frac{\pi^{2}}{12} \frac{m^{2}(h)}{T^{2}}-\frac{\pi}{6}\left(\frac{m^{2}(h)}{T^{2}}\right)^{3 / 2} + \ \mathcal{O}\left[ \frac{m^4}{T^4}\log\left(\frac{m^2}{T^2}\right) \right]~,\notag\\
J_{F}\left(m^{2}(h) / T^{2}\right)&=\frac{7 \pi^{4}}{360}-\frac{\pi^{2}}{24} \frac{m^{2}(h)}{T^{2}}  + \mathcal{O}\left[ \frac{m^4}{T^4}\log\left(\frac{m^2}{T^2}\right) \right]~.
\end{align}

The meson masses in the confined phase are calculated as described in the previous section. We find that for the values of $\xi$ under consideration, the mesons containing top quarks are
typically much heavier than the temperature during the period of early confinement such that they are Boltzmann suppressed. Hence the dominant thermal corrections are from the mesons containing bottom quarks. We keep all 35 mesons in our numerical calculations.

\begin{figure}[t]
\centering
\includegraphics[scale=.6]{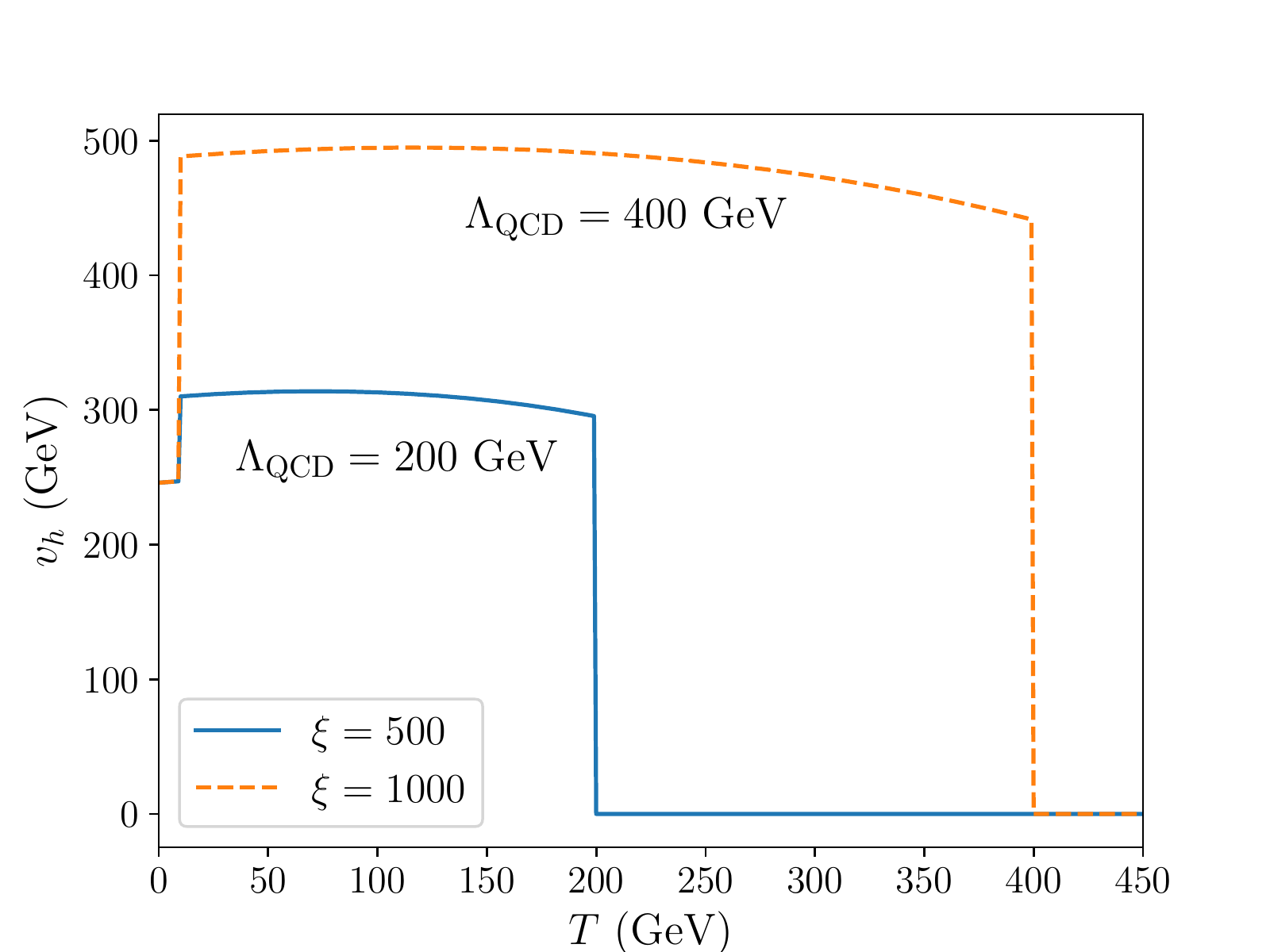}
\caption{Higgs VEV as a function of temperature $T$ for $\xi=500, 1000$ and $T_d = 10$~GeV.
The sudden changes occur at $T \simeq \Lnew$ and $T_d$.
}\label{fig:vevT}
\end{figure}

At high temperatures, the potential is dominated by the $T^2 h^2$ term, driving $v_h \rightarrow 0$, and the
electroweak symmetry is restored.
At $T=\Lnew$, chiral symmetry is broken via the quark condensate, and the tadpole triggers a non-zero Higgs VEV
that is  larger than $v_h^0$ for the $\xi$ values we consider.
At $T_d$, QCD deconfines and the Higgs VEV relaxes to its SM value. This behavior is shown in Figure~\ref{fig:vevT}
for $T_d = 10$~GeV and two values of $\xi$.

\subsection{Dark Matter Interactions with pions}

At leading order in chiral perturbation theory, the interactions with the dark matter map onto,
\begin{equation}
       \frac{\kappa}{M_S^2} ~ \bar{\chi} \chi ~\tr \left( U^\dag \beta + U \beta^\dag \right)
     +
     \frac{i}{M_V^2} ~ \bar{\chi} \gamma^\mu \chi ~\tr \left(  (\partial_\mu U^\dag) ~[ \lambda, U ]
     - [U^\dag, \lambda^\dag] ~ (\partial_\mu U) \right)~,
\end{equation}
with $\kappa$ and $f_\pi$ determined as discussed in Section~\ref{sec:chipt}.  Note that because the scalar interaction with dark matter
is chosen to take the
same form as the spurion containing the quark masses, a single hadronic coefficient $\kappa$ determines both the
pion masses and the dark matter couplings \cite{Bai:2011jg}.
Expanding $U$ to second order for Hermitian choices of $\beta$ and $\lambda$ produces:
\begin{align}
\frac{2\kappa ~ \tr \left[ \beta \right]}{M_S^2} ~ \bar{\chi} \chi +
 \frac{2\kappa}{f_\pi^2}  \frac{1}{M_S^2} ~ \tr[T^aT^b \beta] ~ \bar{\chi} \chi \pi^a \pi^b +
    \frac{2 i}{M_V^2} f^{abc} ~ \tr[T^b \lambda]~ \bar{\chi} \gamma^\mu \chi ~ \pi^a (\partial_\mu \pi^c)  \,.
    \label{eq:interactions}
\end{align}
It is worth noting that the strength of the scalar interaction scales as $\kappa/f_\pi^2 \propto \xi$, whereas the vector-interaction strength is independent of it.

The first term in Equation~(\ref{eq:interactions}) represents a contribution to the dark matter mass induced by the chiral condensate.
At the time of freeze out, the effective mass is given by the sum of $m^{T=0}_\chi$, which to good approximation is
$m_\chi$ in the Lagrangian (\ref{eq:LDM}), and this additional correction that is operative during confinement,
\begin{equation}
    m_\chi^{T=T_F} = m_\chi^{T=0} + \Delta m_\chi~, ~~~{\rm where}~~\Delta m_\chi \simeq (2~{\rm eV})\,\xi^3\left(\frac{10^6~{\rm GeV}}{M_S}\right)^2~. \label{eqn:mshift}
\end{equation}
For large values of
$\xi$, the effective shift may be a few GeV, and may play a role in determining the relic abundance for dark masses of $O(10~{\rm GeV})$. In \Cref{sec:freezeout}  we present our results in terms of the
$T=0$ (unshifted) mass relevant for WIMP searches today.  For dark matter masses of $O({\rm GeV})$, the sign of the effective mass term may flip between
the time of freeze out and today due to a sign difference between $m_\chi$ and $\beta$. For sufficiently complicated WIMP interactions, this could lead to non-trivial interference effects,
but for the simple cases we consider here it is unimportant.

\section{Dark Matter Parameter Space}
\label{sec:freezeout}

In this section, we consider dark matter freezing out through either the scalar or vector interactions introduced above during an early cosmological period of QCD confinement.
We contrast with the expectations from a standard cosmology and constraints from direct searches.

\subsection{Relic Density}
\label{sec:relicdensity}

The Boltzmann equation describing the evolution of the density of dark matter in an expanding Universe can be written as \cite{Kolb:1990vq}:
\begin{equation}
    \frac{dn_\chi}{dt} + 3Hn_\chi = - \langle \sigma v \rangle (n_\chi^2 - n_{eq}^2)\,,
\end{equation}
where $n_\chi$ is the co-moving number density of the dark matter, and $n_{eq}$ is its equilibrium density at a given temperature.
When the interaction rate drops below the expansion rate of the Universe, $H$, the dark matter number density stabilizes,
leaving a relic of the species in the Universe today. The relic density can be solved for a non-relativistic species with a thermally averaged
cross section approximated as
$\langle \sigma v \rangle \sim a + 6 b/x$ where $x \equiv m_\chi/T$. The resulting relic density is:
\begin{equation}
    \Omega_\chi h^2 \approx \frac{1.04 \times 10^9}{M_{Pl}}\frac{x_F}{\sqrt{g_*}} \frac{1}{a + 3b/x_F}\,, \label{eqn:relicdens}
\end{equation}
where $g_*$ counts the number of relativistic degrees of freedom at freeze-out and $h$ parameterizes the Hubble scale. For the standard case of $\xi = 1$, we have $g_*=92$. In an era of QCD confinement at $T\sim 10-100$~GeV, the degrees of freedom changes from the standard scenario since quarks and gluons confine into (heavy) mesons. For the cases we study, this corresponds to $g_*\simeq26$ at the time of dark matter freeze-out.
The freeze out temperature $x_F = m_\chi/T_F$ can be solved for iteratively via
\begin{equation}
    x_F = \text{ln}\left (c(c+2)\sqrt{\frac{45}{8}} \frac{g_\chi}{2\pi^3} \frac{m_\chi M_{Pl}(a+ 6b/x_F)}{\sqrt{g_* x_F}} \right )~, \label{eqn:xfo}
\end{equation}
where $g_\chi=2$ for fermionic dark matter and $c=1/2$ approximates the numerical solution well \cite{Kolb:1990vq}. The parameters $a,b$ in the annihilation cross section are model dependent.  We compute them in \Cref{sec:scalarresults,sec:vectorresults} for scalar and vector interactions, respectively.

The preceding discussion assumes that the freeze out takes place during a time of radiation domination, as is the case for a WIMP in the backdrop of a standard cosmology.
It is generally expected that QCD confinement results in a shift in the vacuum energy of $c_0 \Lnew^4$, where $c_0$ is a dimensionless constant which naive dimensional
analysis would suggest is order 1.  The relic density in Equation~(\ref{eqn:relicdens}) assumes that
the subsequent deconfinement of QCD occurs before the onset of vacuum domination,
\begin{equation}
\Lnew \gtrsim T_F \gtrsim \Lnew \left(\frac{c_0}{g_\ast}\right)^{1/4} .
\end{equation}
For $c_0 \sim 1$, this is a relatively narrow range which would involve some fine-tuning between the freeze out temperature and $\Lnew$ for Equation~(\ref{eqn:relicdens})
to hold.  However, the tiny value of the vacuum energy inferred from cosmic acceleration in the current era could argue that there
is some mechanism at work which dynamically cancels the influence of vacuum energy in different epochs, which would allow for a much wider period of radiation
domination.

\subsection{Limits from Direct Searches}
\label{sec:limits}
Direct detection experiments such as XENON provide important bounds on parameter space based on the null results
for dark matter scattering with nuclei.  The rate for $\chi$ to scatter with a nucleus $N$ in the non-relativistic limit is,
\begin{equation}
    \sigma_{\chi N} = \frac{1}{\pi} \frac{m_\chi^2 m_N^2}{(m_\chi + m_N)^2}[Z f_p + (A-Z) f_n]^2\,,
\end{equation}
where $Z$ and $A$ are the atomic number and mass number respectively and $f_{p/n}$ are the effective couplings
to protons/neutrons, given by
\begin{align}
{\rm Scalar ~Interaction:}&~f_{p/n} = \frac{1}{M_S^2} \left\{ \sum_{q=u,d,s} f_{Tq}^{(p/n)} + \frac{2}{9} f_{Tg}^{(p/n)} \right\}\, , \notag\\
{\rm Vector~Interaction:}&~f_p  = \frac{1}{M_V^2}(3 + \alpha) \,, \quad f_n = \frac{1}{M_V^2}(3 + 2\alpha)\, ,
\end{align}
at leading order \cite{Shifman:1978zn}, with hadronic matrix elements $f_{Tq}$, and $f_{Tg}$
defined as in references \cite{Hill:2014yxa,Mohan:2019zrk}.

\subsection{Scalar-Mediator Results}
\label{sec:scalarresults}

\begin{figure}[t!]
    \centering
    \includegraphics[width=.495\textwidth]{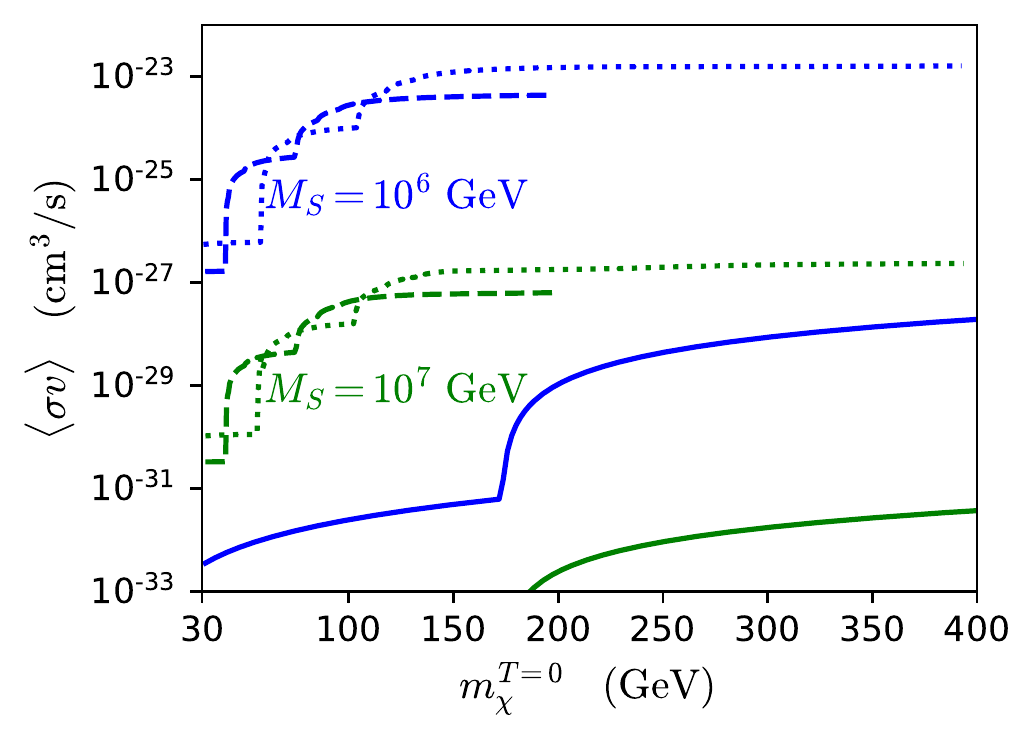}
     \includegraphics[width=.495\textwidth]{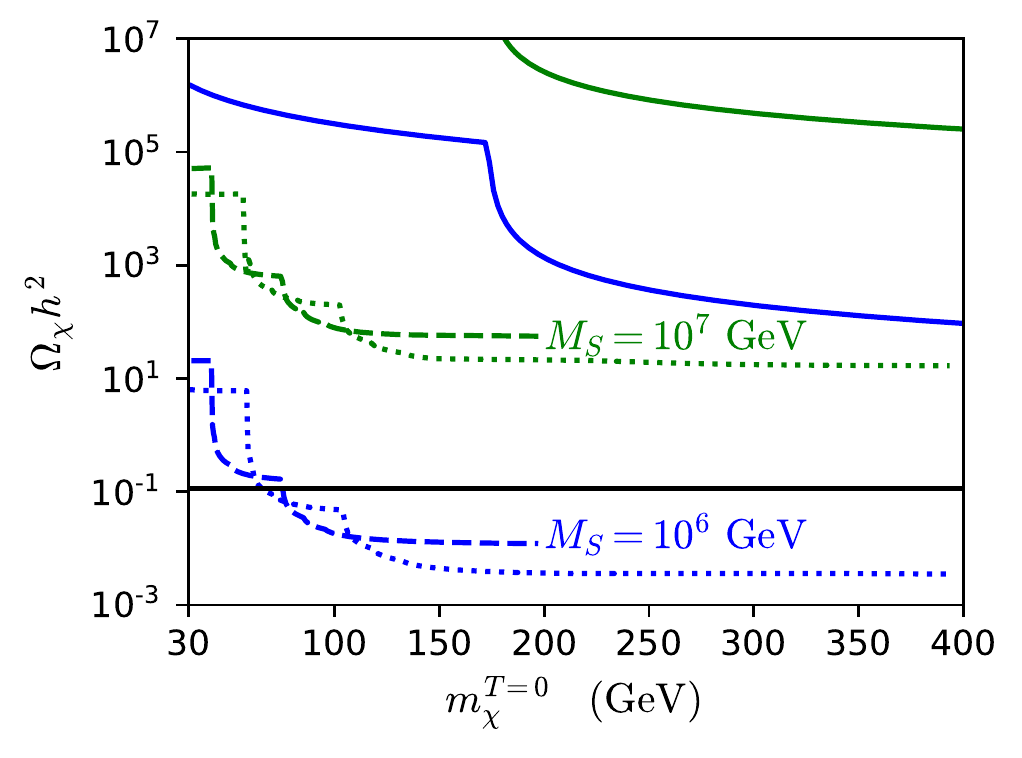}
     \includegraphics[width=.495\textwidth]{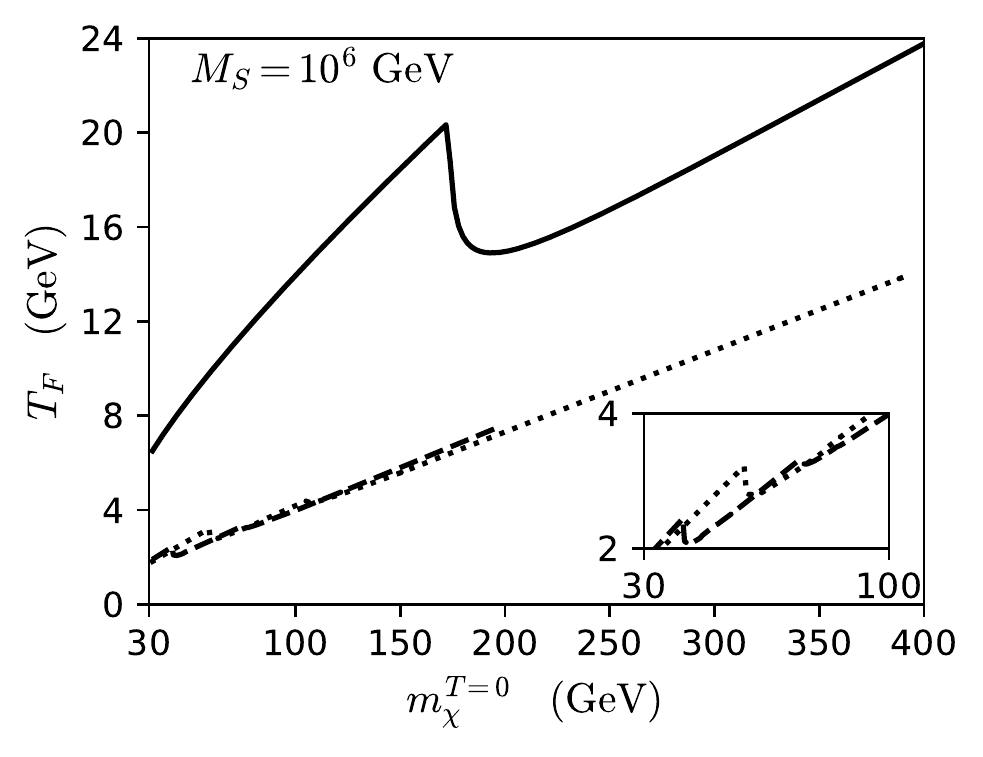}
     \includegraphics[width=.495\textwidth]{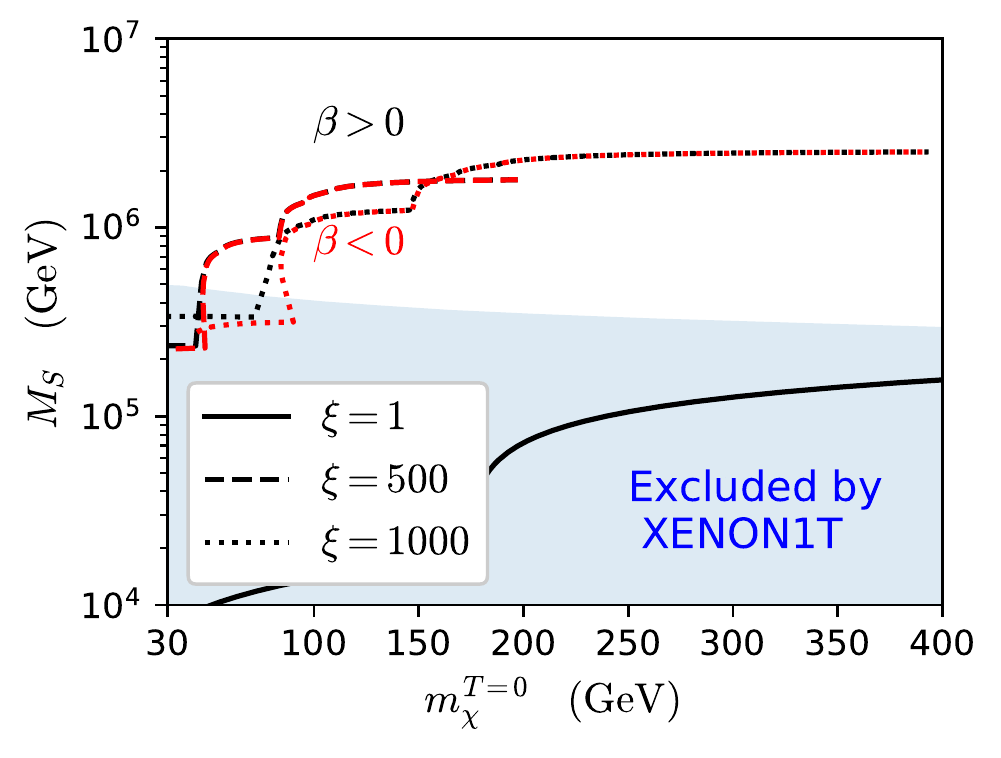}
    \caption{\textbf{(Top Left)} The thermally-averaged cross-sections at the time of freeze-out as a function of $m_\chi^{T=0}$ plotted for $M_S=10^6~{\rm GeV}$ (blue), $10^7$~GeV (green) and $\xi=$1 (solid), 500 (dashed), 1000 (dotted). \textbf{(Top Right)} Dark matter relic abundance today as a function of $m_\chi^{T=0}$ plotted for $M_S=10^6~{\rm GeV}$ (blue), $10^7$~GeV (green) and $\xi=1,500,1000$. The horizontal solid line is the observed dark matter abundance. \textbf{(Bottom Left)} The freeze-out temperature $T_F$ as a function of $m_\chi^{T=0}$ with $M_S=10^6~{\rm GeV},10^7$~GeV plotted for $\xi=$1 (solid), 500 (dashed), 1000 (dotted).  \textbf{(Bottom Right)} We show the $M_S$ values that produce the observed dark matter relic abundance as a function of $m_\chi^{T=0}$ for $\xi=$1 (solid), 500 (dashed), 1000 (dotted). For $\beta < 0$, the line is plotted in red. Shaded blue region is excluded by XENON1T. See text for details. }
    \label{fig:scal_results}
\end{figure}

It can be seen from \Cref{eq:interactions} that the strength of scalar interaction between dark matter and pions depend on the QCD confinement scale, $\Lnew = \xi \LSM$. Consequently, for dark matter with purely scalar interactions, the relic density is a function of the mediator scale $M_S$, QCD confinement scale $\Lnew$, and
the mass of the dark matter at zero temperature, $m_\chi^{T=0}$. We consider
$\xi = 1, 500, 1000$, where $\xi =1$ represents the standard cosmological history and the other two choices
correspond to $\Lnew = 200, 400$~GeV, respectively.

The relic abundance is controlled by the thermally-averaged annihilation cross section at the time of freeze out ($T = T_F$)
in the non-relativistic limit,
\begin{equation}
    \langle \sigma_S v \rangle
    = \left( \frac{\kappa}{f_\pi^2 M_S^2} \right)^2
    \sum_{\pi_a} ~ \frac{\omega_a^2}{4\pi} ~ \sqrt{1-\frac{m_{\pi_a}^2}{m_\chi^2}}\, \langle v^2\rangle + \mathcal O (\langle v^4 \rangle)~. \label{eqn:scal_crossec}
\end{equation}
Here $\omega_a$ are the eigenvalues of the $35 \times 35$ matrix $\tr(T^aT^b \beta)$,
and the sum is over all the pions of mass\footnote{Our choice of couplings $\beta_{ij}$ aligned with the
Yukawa interactions leads to diagonal interactions between the dark matter and the pion mass eigenstates.}
less than $m_\chi^{T=T_F}$.  Note that scalar interactions
lead to $p$-wave suppressed annihilation, for which $a=0$.
The relic abundance today is given by $\rho_\chi = m_\chi^{T=0} n_\chi$,
whereas the energy density immediately after freeze out is $m_\chi^{T=T_F} n_\chi$.
The shift in $m_\chi$ between the time of
freeze out and the present epoch introduces an additional correction to the relic density today:
\begin{equation}
    \Omega_\chi^{T=0} h^2 = \frac{m_\chi^{T=0} }{m_\chi^{T=0} + \Delta m_\chi} \Omega_\chi^{T=T_F} h^2 .
\end{equation}

In Figure \ref{fig:scal_results}, we show the annihilation cross section, relic density today, and freeze out
temperature,
for  $\xi=1,500,1000$ and two representative
values of $M_S$, as a function of the dark matter mass today. In the final panel, we show the value of $M_S$ for each dark matter mass (today) required to reproduce
the observed relic density, for the same values of $\xi$ considered.  Also plotted on that panel are the current XENON1T constraints \cite{Aprile:2018dbl}.  Comparing  $\xi=1$, the standard cosmological scenario, to $\xi=500,1000$ cases makes it clear that freeze-out
during an early cosmological period of QCD confinement, which can realize the observed relic density for weaker couplings, can make the difference between a freeze-out relic WIMP being allowed versus strongly excluded by direct searches.

There are a number of features in Figure \ref{fig:scal_results} that warrant further discussion:
\begin{itemize}
    \item The $\xi \gg 1$ lines end when $m_\chi^{T=T_F}\sim \Lnew \equiv \xi\LSM$, at which point the dark matter mass is heavier than the QCD scale, and the resulting annihilation would be into quarks and not into pions.
    \item For standard cosmology, with $\xi =1$, the kink in the annihilation cross section at $m_\chi \sim 173$~GeV corresponds to the annihilation channel into top quarks opening up.  Similarly, the kinks in the $\xi =500,1000$ lines correspond to new channels into heavier pions.
    \item As mentioned earlier, the annihilation cross section is enhanced by the QCD scale.
    Therefore this scenario accommodates larger values of the mediator scale, $M_S\sim 10^6$~GeV, compared to a standard WIMP scenario.
    \item The effect of the quark-condensate contribution to the dark matter mass can be seen in the bottom-right panel. Depending on the sign of $\beta$ in \ref{eqn:mshift},
    there are two values of $m_\chi^{T=T_F}$ which correspond to a single $m_\chi^{T=0}$ for modest dark matter masses.
    \item The bottom left panel implies that a scenario in which the QCD deconfinement brings the dark matter back into equilibrium with quarks after it has frozen out from interacting with mesons is never realized, for deconfinement happening below a few GeV.
\end{itemize}

\subsection{Vector-Mediator Results}
\label{sec:vectorresults}

\begin{figure}[t!]
    \centering
     \includegraphics[width=.495\textwidth]{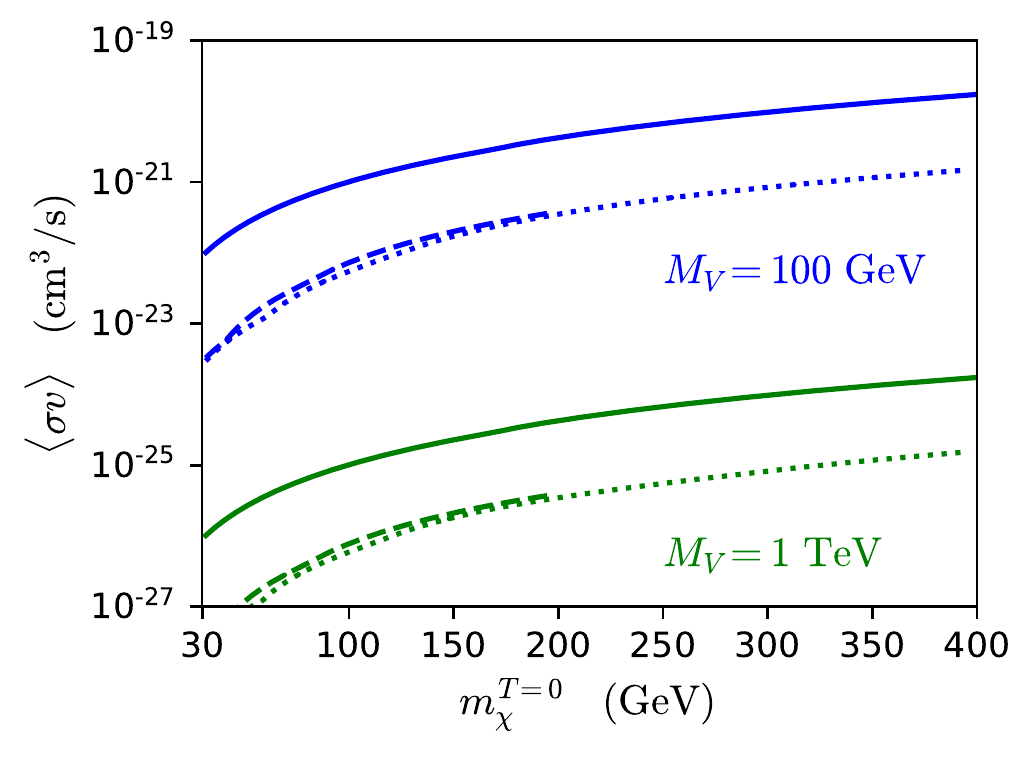}
     \includegraphics[width=.495\textwidth]{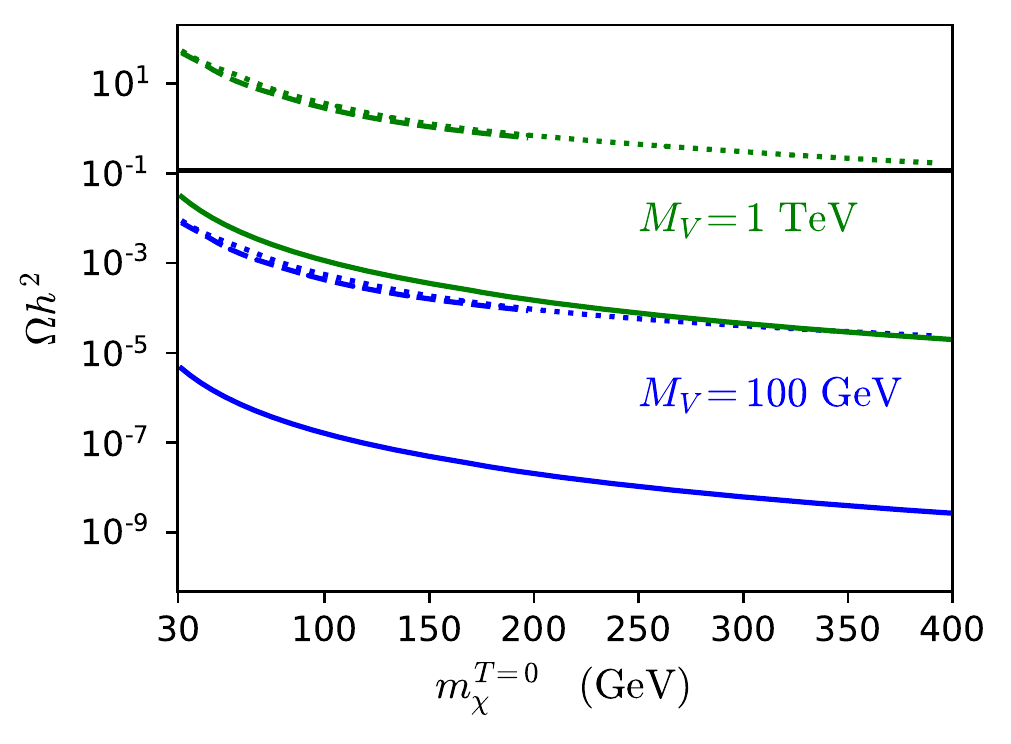}
     \includegraphics[width=.495\textwidth]{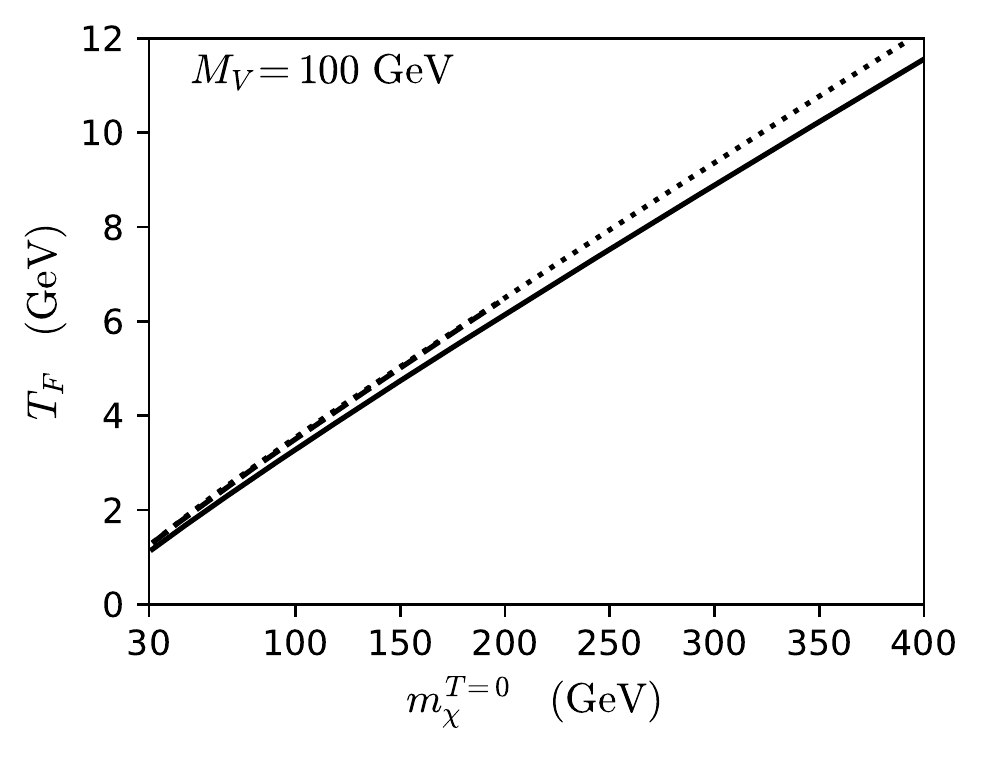}
     \includegraphics[width=.495\textwidth]{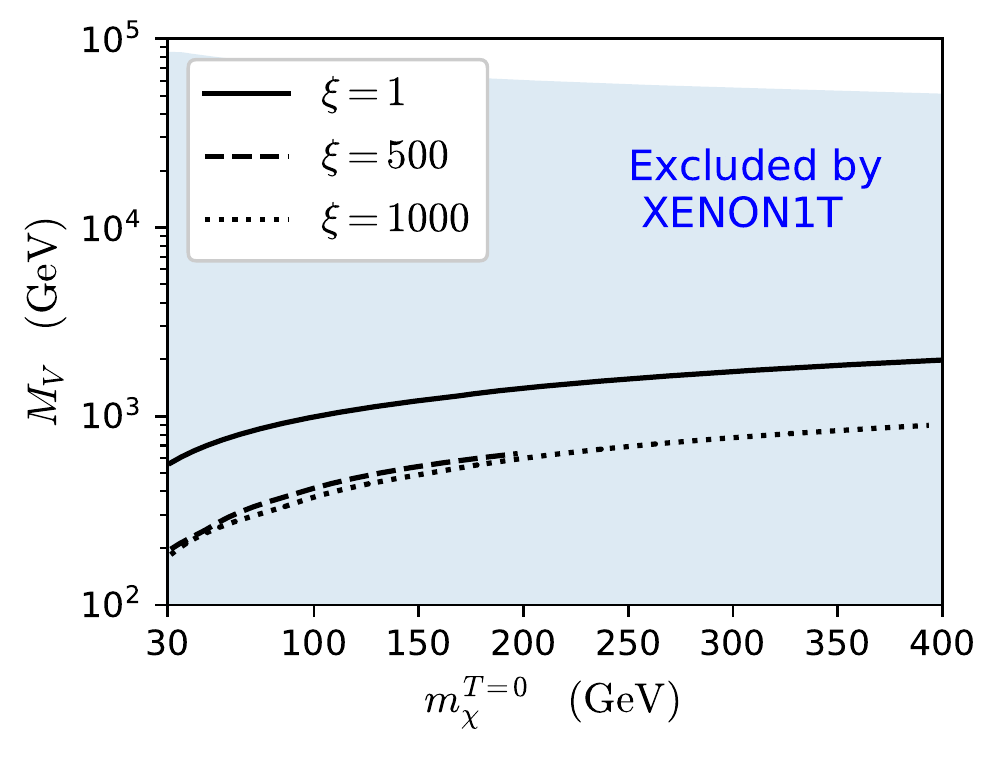}
    \caption{\textbf{(Top Left)} The thermally-averaged cross-sections at the time of freeze-out as a function of $m_\chi^{T=0}$ plotted for $M_V=100~{\rm GeV}$ (blue), $1$~TeV (green) and $\xi=$1 (solid), 500 (dashed), 1000 (dotted). \textbf{(Top Right)} The generated relic abundance today as a function of $m_\chi^{T=0}$ plotted for $M_V=100~{\rm GeV}$ (blue), 1~TeV (green) and $\xi=$1 (solid), 500 (dashed), 1000 (dotted). The horizontal solid line is the observed dark matter abundance. \textbf{(Bottom Left)} The freeze-out temperature as a function of $m_\chi^{T=0}$ with $M_V=100~{\rm GeV}$ plotted for $\xi=$1 (solid), 500 (dashed), 1000 (dotted). \textbf{(Bottom Right)} Coupling as a function of $m_\chi^{T=0}$ to produce the observed relic density plotted for $\xi=$1 (solid), 500 (dashed), 1000 (dotted). Shaded blue region is excluded by XENON1T. See text for details.}
    \label{fig:vec_results}
\end{figure}

For vector interactions, our choice of minimally flavor-violating interactions $\lambda_{ij}$ with the quarks results in leading interactions
with a pair of pions, as in Equation~(\ref{eq:interactions}).  In the non-relativistic limit, the thermally-averaged annihilation cross section is,
\begin{equation}
   \langle \sigma_V v \rangle= \sum_{a,b=1}^{35} \frac{\Omega_{ab}}{24\pi}m_\chi^2(1-\gamma_{ab} + \rho_{ab})^{3/2} \left [ 1 + \left (1 + \frac{9}{4}\frac{\gamma_{ab} - 2\rho_{ab}}{1-\gamma_{ab} + \rho_{ab}} \right )\frac{\langle v^2 \rangle }{2} + \mathcal O (\langle v^4 \rangle )\right  ] \label{eqn:vec_crossec}
\end{equation}
summed over pairs of mesons for which $m_{\pi_a} + m_{\pi_b} \leq 2 m_{\chi}$.  Note that vector interactions do not induce a shift in the mass of the dark matter
from the chiral condensate.  The coupling matrix $\Omega_{ab}$ is given by
\begin{equation}
    \Omega_{ab} \equiv \sum_{c,d=1}^{35} \frac{1}{M_V^4}f^{abc}\,\tr[T^c \lambda] \,f^{abd}\,\tr[T^d \lambda] \propto \frac{\alpha^2}{M_V^4}~,
\end{equation}
where we focus on $\alpha=1$ for simplicity.
The kinematic factors are defined as $\gamma_{ab} \equiv (m_{\pi_a}^2 + m_{\pi_b}^2) / (2 m_\chi^2)$
and $\rho_{ab} \equiv (m_{\pi_a}^2-m_{\pi_b}^2)^2 / (16m_\chi^4)$.

In Figure~\ref{fig:vec_results} we show the resulting annihilation cross section, relic density, and freeze-out temperature, as a function of
the dark matter mass at zero temperature $m_\chi^{T=0}$, for two choices of $M_V=100$ GeV, 1 TeV and $\xi = 1,500,1000$, where $\xi=1$ corresponds
to the standard picture of freeze-out through annihilation into quarks.  In the final panel, we show the value of $M_V$ for each dark matter mass required
to reproduce the observed relic density for a given choice of $\xi$.

Unlike the scalar interactions, vector interactions do not get the $\xi-$enhancement from QCD confinement. On the contrary the annihilation cross-section is smaller than the standard WIMP scenario because the annihilation products, namely the new pions, are heavier than SM quarks at the same temperature in standard cosmology. Hence, the vector scenario does worse than the standard WIMP case within this cosmological history.

\section{Conclusions}
\label{sec:conclusions}

The standard picture of freeze out is a compelling picture for the mechanism by which dark matter is produced in the early Universe, and the primary
motivation for WIMP dark matter.  Common wisdom states that the WIMP paradigm is in trouble, but this is the result of comparing freeze out in a standard cosmology
to searches for WIMPs.  In this article, we have explored the possibility that the cosmology looks radically different at the time of freeze out, in particular exploring the
idea that QCD could have undergone an early period of confinement before relaxing to the behavior observed at low temperatures today.  We find that for a scalar
mediator, the dark matter mass is shifted by the chiral condensate, and its coupling to pions is enhanced during early confinement, allowing for parameter space
which allows for freeze out production while remaining safe from constraints from XENON1T today, rescuing some of the WIMP parameter space.  On the other hand,
for a vector mediator we find that the differences between freeze out during early confinement and the standard cosmology are more modest, and the entire parameter
space remains ruled out by XENON1T.  Our work highlights the fact that a modified cosmology may largely distort the apparent messages from astrophysical observations
of dark matter to inform particle physics model building.

\section*{Acknowledgements}
The authors are grateful for discussions with C.~Csaki, M.~Geller, J.~Unwin and M.~Luty.
This work was supported in part by the NSF via grant number PHY-1915005 and DGE-1839285.  SI acknowledges support from the University Office of the President via a UC Presidential Postdoctoral fellowship.

\bibliographystyle{JHEP}
\bibliography{references}
\end{document}